# Digital Resistance during COVID-19: A Workflow Management System of Contactless Purchasing and Its Empirical Study of Customer Acceptance

Yang Lu[1]

Information Systems and Operations Management, College of Business, University of Central Oklahoma, Edmond, OK 73034

[1]ziiyuu@gmail.com

**Digital Resistance during COVID-19: A Workflow Management System of Contactless Purchasing and Its Empirical Study of Customer Acceptance**

Abstract

The COVID-19 pandemic has stimulated the shift of work and life from the physical to a more digital format. To survive and thrive, companies have integrated more digital-enabled elements into their businesses to facilitate resilience, by avoiding potential close physical contact. Following Design Science Research Methodology (DSRM), this paper builds a workflow management system for contactless digital resilience when customers are purchasing in a store.

The findings show that response costs have a positively significant effect on customers' behavioral intention to adopt digital resilience, while self-efficacy plays a negative role on customers' behavioral intention. The findings reveal that, during the COVID-19 pandemic, customers are more concerned about health issues and put more effort into the deployment of digital resilience to mitigate the consequences of the virus. These results indicate that, even beyond the performance of technology itself, another factor (the health issue) can play the key role in customers' acceptance of digital resilience.

# INTRODUCTION

Until the end of September 2020, the cumulative number of confirmed cases of COVID-19, worldwide, stood at 34.1 million. The number of deaths, at that point in time, was 1.02 million[1]. The COVID-19 pandemic has caused dramatic damage and has thoroughly changed organizations' operating modes, as well as people's lives and habits. Close physical contact is the major reason given for the fast spread and infection of COVID-19 (Guan et al., 2020; Wu & McGoogan, 2020; Sohrabi et al., 2020). As one effective measure, digital resilience is being deployed, and it is quickly being improved to its highest level, in order to mitigate the influence of the pandemic. Both companies and individuals hold virtual meetings instead of face-to-face ones. Since the inception of COVID-19, the popular virtual communication and conference software called ZOOM's stock price has skyrocketed to the price of $559 on October 16, 2020, up from around $70 in January 2020.

Close physical contact[2] happens among people every day; contact is unavoidable. Walmart provides three types of grocery shopping: purchasing at a local store (conventional), ordering online with pickup (blended), and ordering online with delivery (e-commerce). It is important to continue to offer the conventional purchasing style, since many customers still prefer to select their food themselves or because certain categories of food are not available when using the other two purchasing styles. Because of these, there is an urgent need for businesses to assist their customers in avoiding potential close physical contact. This study focuses on the first purchasing style, conventional purchasing, by designing a digital resilience workflow management system that helps customers avoid potential close physical contacts when they are purchasing in a store. This robust and applicable digital infrastructure will enhance companies' resilience in fighting against the virus and will assist in making more profitable businesses, as well.

---

[1] Source: (WHO) https://covid19.who.int, (CDC) https://www.cdc.gov/coronavirus/2019-ncov/index.html, and (Johns Hopkins Coronavirus Resource Center) https://coronavirus.jhu.edu/map.html. Accessed on September 30, 2020.

[2] Close physical contacts refer to contacts that are within 2 meters (6 feet) for over 15 minutes.

As the groundwork of information systems, digital resilience describes an organization's capability to deal with unexpected disruptions, in order to continue doing business and to be successful after the emergency. Digital resilience has the potential to change not only an organization's operating modes, but also people's behavior and habits. When they are trying to mitigate the influences of the COVID-19 pandemic, companies, for their businesses to benefit, need to deploy more resiliency-related digital techniques; individual customers, for their health, need to cope with these resiliency-related digital strategies to avoid potential infection. Information systems is an important discipline that can be used to explore insights that can help to resolve the many issues caused by the unexpected COVID-19 disruptions. Digital resilience can be achieved through information systems' integration of high technology with advanced devices. A well-designed digital resilience workflow management system can sustain businesses' continuity and can mitigate the impact of COVID-19. Our research lies mainly in helping information systems to find a way to accelerate digital resilience during the COVID-19 period.

This study articulates three objectives:

First, it should be noted that design science is a distinguished and classical methodology among the IS disciplines. This paper uses the framework of Design Science Research Methodology (DSRM) not only to identify the critical problem of potential close physical contact in the COVID world, but to define the objectives of the proposed workflow management system by flowchart using the epidemiological SIR model, to design and to illustrate the digital resilience flow of contactless purchasing by using the Petri net workflow management system, and to assess the feasibility of the workflow management system by using behavioral theories associated with empirical study. DSRM is considered an interdisciplinary methodology composed of design science and empirical research, in this paper.

Second, the workflow management system described herein is built to embed digital resilience, in order to allow businesses, the chance to help their customers avoid potential close physical contact when they are making a purchase in a store. Digital resilience is a must-have tool for a business' continuity and performance, especially in the event of an emergency. The proposed workflow system offers good guidance that a company can follow, so that it can continue to be successful despite unexpected disruptions – having previously invested in digital resilience and having facilitated digital resilience into its enterprise management system.

Last but not least, depending on the workflow management system of contactless purchasing, the feasibility of the proposed workflow management system should be considered. This workflow depicts the human behaviors of considering, recognizing, coping, behaving, and using digital resilience to prevent potential close physical contacts under the COVID-19 pandemic. If customers are reluctant (or are not able) to adopt digital resilience measures when making a purchase in a store, the company can still effectively change and successfully implement digital resilience to keep its customers away from potential infection. This paper describes an empirical examination that considers what factors most relate to customers' intention to cope with digital resilience in a store, and we found two factors that have different impacts on customers' intention. The response cost (facilitating conditions) is positively associated with customers' digital resilience adoption, and self-efficacy (facilitating conditions) is negatively associated with customers' digital resilience adoption. COVID-19 has changed people's behavior and habits, not only when making a purchase in a store but also when accomplishing many other daily activities, e.g., wearing masks and hand sanitizing.

The overall infrastructure of this study (**Appendix Figure A1**) follows the main steps of Design Science Research Methodology (DSRM) (Hevner et al., 2004; Peffers et al., 2007; Vandenbosch & Higgins, 1995; Carvalho, 2020): (1) problem identification and motivation, (2) definition of the objectives for a solution, (3) design and development of the model, (4) demonstration of the model, and (5) evaluation of the model. We use DSRM to propose a Petri net workflow management system that guides customers away from potential close physical contact when they are making a purchase in a store.

## PROBLEM IDENTIFICATION AND MOTIVATION OF DIGITAL RESILIENCE

The first step of DSRM is to identify problem and motivation. The Novel COVID-19 is a coronavirus that is similar to SARS-CoV and MERS-CoV (Lai et al., 2020; Shereen et al., 2020; He, Deng, & Li, 2020). The difference is that COVID-19 spread all across the globe, as a pandemic, within a short six-month period, causing huge impacts on people's lives and on society (Li et al., 2020; Chan et al., 2020). In this paper, the epidemiological SIR model is employed to explore the reason why so many people have become infected. The main reason appears to be the close physical contact between infected (symptomatic and asymptomatic) and susceptible people.

### The SIR Model

In general, the SIR model consists of three compartments (**Figure 1.**), susceptible (*S*), infectious (*I*), and removed (*R*). *S* describes the people who are susceptible to the disease. At the beginning of the pandemic, *S* equals to the total population in a certain area. *I* describes the people who are infectious. The infectious people have the disease, and they can infect others. *R* (or removed) describes the people who have caught the disease and who have now either recovered from it or have died. These recovered people are immune to the disease. Thus, the removed people are those who are not infectious anymore (Chen et al., 2020; Lin et al., 2020).

In the SIR model, several assumptions are used to simplify the real-world phenomenon of COVID-19. Explanations of the variables of SIR model are addressed in **Appendix 2. (Table A1)**.

(1) The total population ($TP$) remains constant during the pandemic. It means that the rate of change of the susceptible population plus the rate of change of the infectious population plus the rate of the removed population must be zero. The total population ($TP$) is given by ($S+I+R$).

$$TP = S + I + R = I_0 + S_0$$

$$d/dt\ (S+I+R) = (-\gamma * I * S) + (\gamma * I * S - \alpha * I) + (\alpha * I) = 0$$

This will be the same constant value for all the possible values of time. The initial value will be the starting point: the value of the total population at the beginning of the pandemic. As time progresses, it will not change. It will always equal the initial value.

(2) The transmission rate ($\gamma$) is proportional to the contact between the susceptible and the infectious people. And $\gamma$ occurs at a constant rate. The transmission rate ($\gamma$) will decrease as more people become infectious.

(3) The removed rate ($\alpha$) is a constant rate. It could be a death rate or a recovery rate, or it could be the composite of the death and recovery rates.

(4) The contact ratio ($q$) is the fraction of the population that comes into contact with an infected individual during the period when they are infectious, $q = \gamma / \alpha$.

(5) The basic reproductive ratio ($R_0$) is the reciprocal of the contact ratio ($q$), $R_0 = \alpha / \gamma$. This ratio indicates that there will be an epidemic if $R_0 > 1$

(6) The initial number of susceptible people is $S_0$, the initial number of infectious people is $I_0$, and the initial value of removed people is 0.

The rate of change of the number of susceptible people over time:

$$dS/dt = - \gamma * I * S \qquad (1)$$

The rate of change of the number of infectious people over time:

$$dI/dt = \gamma * I * S - \alpha * I \qquad (2)$$

The rate of change of the number of removed people over time:

$$dR/dt = \alpha * I \qquad (3)$$

These three differential equations are for the three compartments of people of the population. Equation (1) indicates that the number of susceptible people is going to change according to the number of contacts between susceptible and infectious people. Equation (2) indicates that the number of infectious will increase because of the contact between people who have either recovered or died as a result of the disease spread. Equation (3) indicates that the rate of removed people is going to increase at the constant rate, depending on how many infectious people there are.

The SIR model assumes that susceptible people will transfer to other states with a certain probability of infection, according to the development pattern of COVID-19. The dynamic model of "susceptible-infectious-removed" can predict the trend of COVID-19 within a certain range, geographical area, or time segment.

**Evaluating the Importance of Contact Ratio ($q$)**

The initial number of susceptible people is $S_0$, the initial number of infectious people is $I_0$, and the initial value of removed people is 0. The following equation is the initial point of COVID-19.

$$S+I+R = I_0 + S_0 \qquad (4)$$

Next, we investigate and discuss three important issues of COVID-19 based on the SIR model: the severe spread, the potential maximum number of infectious people, and the potential number of infected people by the end of the pandemic. All three problems are related to the contact ratio ($q$).

**The Severe Spread of COVID-19**

The initial number of infectious people at the beginning of the outbreak is given by $I_0$. The question is whether or not the number of infectious people will grow. If the number of infectious people starts to grow, the disease will spread throughout the population. Here, we focus on Equation (2), the rate of change of infectious people over time. $S$ is smaller than its initial value ($S \leq S_0$). In the context of the disease, at the beginning of the outbreak, everyone in the total population theoretically was susceptible to the disease, especially since it was a Novel Coronavirus, i.e., one that had never been seen before.

Since $S \leq S_0$, we have

$$dI/dt < I\,(\gamma * S_0 - \alpha) \qquad (5)$$

An epidemic will occur if the size of *I* increases from the initial value of infectious people ($I_0$). In the very real situation of COVID-19, it became clear that the number of infectious people was increasing very quickly. For the other part of Equation ($\gamma S_0 - a$), if this term is positive, there will be a spread of the disease. It means,

$$S_0 > \alpha / \gamma \qquad (6)$$

The basic reproductive ratio $R_0 = \alpha / \gamma$. This ratio indicates that there will be an epidemic if $R_0 > 1$. This ratio represents the secondary infections in the population caused by one initial primary infection. In other words, if one person has the disease, $R_0$ will show how many infections, on average, that person is likely to cause. This current coronavirus is an ongoing outbreak that we have never seen before. The reproductive ratio, as described in the research, is estimated to be more likely 2 to 4 (Chen et al., 2020; Li et al., 2020). COVID-19 is an epidemic that spreads quickly. Therefore, avoiding potential close physical contact is an effective way to reduce the contact ratio and to decrease the number of infected people.

**The Potential Maximum Number of Infectious of COVID-19**

There is a known lack of appropriate and effective approaches to detection and diagnosis in the early stages of any disease outbreak, especially in unknown epidemics like COVID-19. Knowledge of the precise estimate of the number of people infected is essential, in order to be able to judge the severity of the epidemic and to make corresponding decisions. A common method used is to estimate the number of infections based on the proportion of outflowing people in a certain area. The early report from Northeastern University (Chinazzi et al., 2020) made a similar relevant analysis.

Knowing the number of infected people is very helpful when it comes to planning how to distribute health resources and how to implement anti-COVID measures. In Equations (1) and (2),

$$dI/dS = (\gamma IS - aI)/(-\gamma IS) = -1 + a/\gamma s \quad (7)$$

The contact ratio $q = \gamma / a$, we have

$$I + S - 1/q * \ln S = I_0 + S_0 - 1/q * \ln S_0 \quad (8)$$

The maximum will occur, when $S = 1/q$. Substituting this value into the equation (8),

$$I_{MAX} = I_0 + S_0 - 1/q \, (1 + \ln(qS_0)) \quad (9)$$

The maximum number of infectious people ($I_{MAX}$) is the maximum number of people that will have the disease at a given time. The term ($1/q \, (1 + \ln(qS_0))$) depends on the parameter $q$, the contact ratio. In the outbreak of COVID-19, the value of $q$ is high; the disease is very easy to transmit. Many susceptible people are becoming infected when encountering potential close physical contact with infectious people, especially since COVID-19 has a relatively long incubation period, during which its symptoms might not yet have appeared. Avoiding potential close physical contact separates the susceptible from the infectious people, in order to reduce the quantity of overall infectious (both symptomatic and asymptomatic) people.

**The Potential Number of Infected People by the End**

How can we know that the pandemic is at its end? The number of infectious people will go down to zero. This, in the future, will signal the end of the outbreak. Let us rearrange to find the size of the removed people (R), those who have either recovered or died, at the end of the pandemic. The total number of people who have caught the disease by the end is,

$$R(end) = I_0 + S_0 - S(end) \quad (10)$$

Based on Equation (8), the removed people or the size of the removed population at the end of the epidemic is,

$$S(end) - 1/q * \ln(S(end)) = I_0 + S_0 - 1/q * \ln(S_0) \quad (11)$$

If the value of $q$ is sufficiently large, most of the population will not catch the disease. In the case of COVID-19, if there is a large value of $q$, the potential maximum number of infectious people at any given time is almost equal to the whole population, in theory.

In summary, the contact ratio ($q$) appears in the answers to all three key questions. It is impossible to stop the spread of COVID-19 that has already occurred; what we can do is reduce the number of people who will get infected ($I_{MAX}$). It is practical to isolate the susceptible people from the infectious people. This is exactly why we need to avoid potential close physical contact.

In reality, grocery shopping has become one of the major channels to explore, during the COVID-19 era. Our study depicts a workflow management system to solve the issue of potential close physical contact when a customer is making a purchase in a store. The proposed workflow management system will contribute to the IS community's fight against the COVID-19 pandemic.

### DEFINITION OF THE OBJECTIVES OF DIGITAL RESILIENCE MEASURES

The second step of DSRM is to interpret the objectives of a solution. Administrative authorities have suggested several policies to be taken against COVID-19, such as staying at home, avoiding gatherings or parties, closing stores and places to shop, etc. However, people cannot escape their need for groceries. There are different groups of personnel at grocery stores; this leads to a complicated COVID-19 infection network fraught with potential close physical contacts. To mitigate infection, a store can deploy anti-COVID measures; digital resilience is one of the most effective ways to avoid potential physical contact.

Our goal is to mitigate the influence of the COVID-19 pandemic, specifically by focusing on avoiding close physical contacts between a business and its customers, by the use of digital resilience. A flowchart (**Figure 2.**) illustrates how a customer can avoid potential physical contact when making a purchase in a store. The detailed procedure and the relevant activities in this digital resilience system are described and explained in the next section.

# DESIGN OF DIGITAL RESILIENCE WORKFLOW MANAGEMENT SYSTEM

## Workflow Management System of Avoiding Contact

A Petri net workflow (Salimifard & Wright, 2001; Xu et al., 2009) is built to help customers avoid close potential physical contact in a store. The proposed workflow management system consists of five major procedures: the Entering Procedure (EP), the Purchasing Procedure (PuP), the Payment Procedure (PaP), the Delivery Procedure (DP), and the Customer Service Procedure (CSP), as well as six role players, the Customer (C), the Sensor Checking System (SC), the Purchasing Monitoring System (PM), the Payment Assistant System (PA), the Delivery Assistant System (DA), and the Customer Service System (CS).

Each role player is represented by a labeled Petri net (LPN) model, and all LPN models are combined as the complete workflow management system. The system includes five interactive transactions: the interactions between C and SC in EP, between C and PM in PuP, between C and PA in PaP, between C and DA in DP, and between C and CS in CSP. Within the entire process, many digital resilience-enabled devices and sensors are available to assist customers. From a customer behavioral perspective, companies can recognize which factors are likely to impact their customers' intention and can adjust accordingly.

## Labeled Petri Net Workflow Management System

The proposed labeled Petri net model is constructed based on previous studies (Van der Aalst, 1998 & 2000; Xu et al., 2009). We constructed a labeled Petri net model (LPN) and a labeled workflow net (LWN). LPN represents each role player, and LWN represents the complete system. Transitions are divided into three categories: In, Out, and Inner transitions. The In Transition refers to "receiving a message from a partner via network"; the Out Transition refers to "sending a message to a partner via network"; and the Inner Transition "contains all inner activities" (Du, Jiang, & Zhou, 2009; Du et al., 2009). In the proposed workflow management system, customer and the five assistant systems are partners, and all messages and relevant activities are interacted between these six role players throughout the system. All messages and activities are recorded in the system for further analysis.

**Definition 1**. A labeled Petri net (LPN) is composed of 7 tuples,

$$LPN = (P, T, F, M_0, \varphi, S_l, F_l).$$

Criteria:

(1) $P$ is a finite set of places.

(2) $T$ is a finite set of transitions. $T = T_{In} \cup T_{Out} \cup T_{Inner}$. The three categories (In, Out, and Inner) are mutually exclusive in a workflow system.

(3) $F \subseteq (P \times T) \cup (T \times P)$, which refers to a set of directed arcs (relations) connecting Places to Transitions and Transitions to Places.

(4) $(P, T, F)$ represents a Petri net.

(5) $M: P \rightarrow \{0,1\}$ is a marking function. $M_0$ is the initiation marking.

(6) $\varphi$ is the set of messages between customers and business. Each message is defined as the form of [(msg, Sender, Receiver)]; msg is the name of a specific message or task.

(7) $(M, \varphi)$ is a state of LPN. $(M_0, \varphi_0)$ is an initial state, where $\varphi_0$ is a non-empty set.

(8) $S_l$ is a finite set of activity labels, e.g., Greek or Arabic.

(9) $F_l: T \rightarrow S_l$ is defined as a labeling or weight function.

**Definition 2.** $LWN = (P, T, F, M_0, \varphi, S_l, F_l) = LPN$.

Labeled workflow net (LWN) is an LPN, if and only if

(1) $P$ consists of a source place $i$, which is a non-empty set.

(2) $P$ consists of outcome places $O_i$, which is a non-empty set.

## DEMONSTRATION OF DIGITAL RESILIENCE WORKFLOW MANAGEMENT SYSTEM

In the proposed LWN, $P$ ($P_1, P_2, \ldots, P_{20}$) is place that is expressed by a circle. The $T_{In}$ and $T_{Out}$ transitions are represented by rectangles with exchanged messages. The $T_{Inner}$ transition is represented by a solid rectangle. The terminal goal is $G = \{M(O_1)=1; M(O_2)=1; M(O_3)=1\}$. Specifically, $M(O_1)=1$ indicates that a customer's access to a store has been denied because the customer has failed a physical temperature check. $M(O_2)=1$ indicates that a customer's access to a store has been denied because the customer has refused to wear a mask. $M(O_3)=1$ indicates that a customer has successfully finished a purchasing process in a store with the assistance of the digital resilience workflow management system, which has provided store access check (the store's customer capacity, the customer's temperature, and the wearing of a mask); purchasing process assistance (crowd density, one-way direction); self-payment system (cash, card, or App Pay); delivery assistance (a self-delivery system); and customer service (a self-customer service system). Messages are exchanged between the customers and the business. (Access, C, B) means that a store receives an access request from a customer; Out (N_Tem, SC, C) means that the Sensor Checking System sends a message of a customer's temperature fail from SC to C. More detailed explanations of the messages are shown in **Appendix** (**Table A2**).

In the proposed workflow management system, there are three terminal goals: $O_1$, $O_2$, and $O_3$. Only $O_3$ consists of all the possible digital resilience-enabled purchasing processes. Both $O_1$ and $O_2$ deny access to a store because of a temperature check failure or no mask wearing, respectively. Let us have a detailed look at $O_3$ from the starting point $i$. The complete workflow system (**Figure 3.**) includes the five procedures mentioned above.

In the first procedure, Access, the interaction between C and SC in EP involves several sensors and protocols. Since the COVID-19 pandemic is severe, every customer is required to follow three access checks: the store capacity check ([(Cap, SC, C)]), the body temperature check ([(Temp, SC, C)]), and the mouth and nose mask check ([(Mask, SC, C)]). Here is the order: first, a customer requests access to a grocery store ([(Access, C, B)]). If that store already has its maximum number of customers inside, as a safety issue, the customer ([(Y_Cap, B, C)]) is told to wait to enter until another customer finishes shopping. If a store does not have its maximum number of customers, the customer ([(N_Cap, B, C)]) will be allowed to enter if the customer satisfies the temperature ([(Y_Tem, SC, C)]) and mask wearing ([(Y_Mas, SC, C)]) requirements. Any customer will be denied entry to a store if the customer is reluctant either a) to check his/her body temperature or b) to wear a mask ([(N_Mas, SC, C)]). If a customer has a temperature check and shows a temperature that is above the normal range, the customer ([(N_Tem, SC, C)]) will be denied access to the store. All three activities would be monitored and controlled by digital devices, with notice and instructions sent to the customer. It is voluntary that customers complete extra anti-infection measures, such as hand sanitizing, cart cleaning, and wearing gloves, etc.

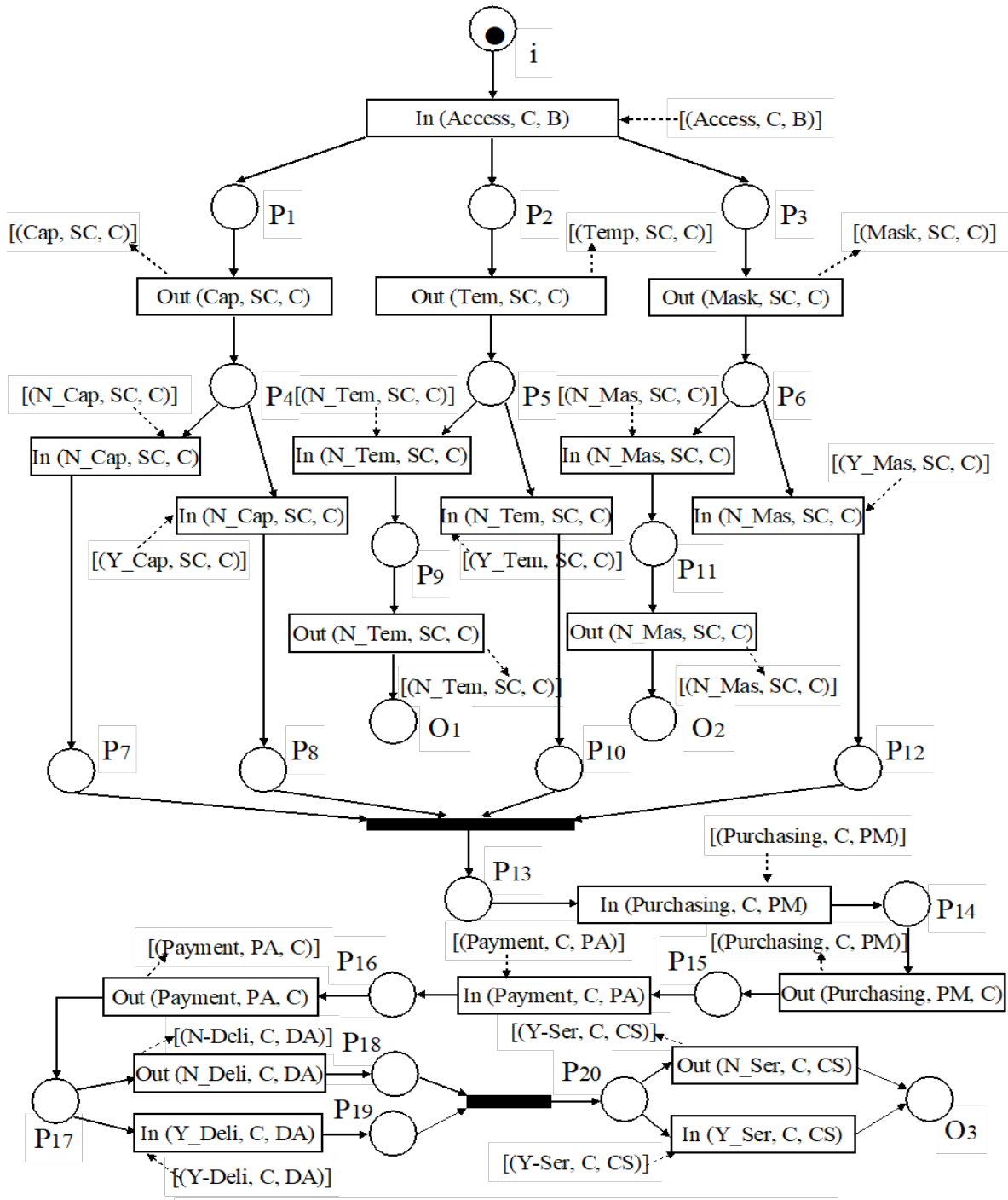

Figure 3. The Workflow Management System of Digital Resilience

In the second procedure, Purchasing, the interaction between C and PM in PuP, digital resilience measures will assist and warn customers ([(Pur, PM, C)]), all throughout the store, if a certain area has a dense crowd or if the customer has not followed the correct direction during shopping. Customers can also install the related App to track and to instantly obtain useful information.

In the third procedure, Payment, the interaction between C and PA in PaP, there is no personal assistant. What the customer ([(Pay, C, PA)]/ [(Pay, PA, C)]) needs to do is adopt a self-assistant system to scan and pay for his/her items by cash or by card. Another potential digital resilience measure is that a customer can use his/her own cell phone to scan and pay through a payment App.

In the fourth procedure, Delivery, the interaction between C and DA in DP, a customer ([(Y_Deli, C, DA)]) can use a digital device to process a delivery if the customer needs some of the items to be delivered. If there is no request from the customer ([(N_Deli, C, DA)]) to deliver anything, the customer will be directed to the final step: Customer Service.

In the fifth procedure, Customer Service, the interaction between C and CS in CSP, many types of contactless service can be implemented, such as a voice assistant, a virtual assistant, an App assistant, a message assistant, etc. If the customer ([(Y_Ser, C, CS)]) needs customer service, the system will assist him/her. If not, the system will finish all of its possible assisting and the customer's purchasing will end at $O_3$.

## THEORETICAL CONTRIBUTIONS

This study is a good attempt to present a workflow management system of digital resilience to mitigate consequences of the COVID-19 pandemic by integrating the two major research paradigms of information systems: design science and behavioral research. First, the structure and the context of this paper are based on DSRM (Design Science Research Methodology). That methodology is suitable to use in identifying a practical problem (the potential close physical contacts of customers during purchasing in a store), in building a workflow management system to help businesses' customers avoid potential close physical contacts, and in empirically evaluating the feasibility of the system (whether or not customers will adopt digital resilience when making a purchase, and what factors impact customers' behavioral intention). DSRM is an appropriate measure, both theoretical and indirect, to use in mitigating the influence of the unexpected disruptions.

## LIMITATIONS AND FUTURE RESEARCH

While our paper focuses on the ways in which customers can accept digital resilience measures taken to counter the risks of COVID-19, another important angle to consider is the employees' prospects for digital resilience. A real-life example happened at Walmart's "Order Online & Pickup." Walmart had already made a digital resilience effort; the App indicated that when a customer arrived to make a pickup, he or she should "Roll Up" the vehicle's windows to protect the driver and the employee from potential infection. However, as happens often, neither the employees nor the customers obeyed this principle, because it was easier to open the window for communication between employees and customers. Nevertheless, during the COVID-19 pandemic, we must learn to tolerate the inconvenience of avoiding potential close physical contact. Employees need to be trained and must follow the policy of digital resilience in order to avoid any potential close physical contact. Businesses must learn how to monitor and manage their employees' behavior regarding digital resilience. Otherwise, digital resilience may not perform well in mitigating COVID-19 influences or other unexpected disruptions.

Previous studies have explored the behavioral model's relationship to moderating effects, such as age, gender, education level, work experience (Compeau & Higgins, 1995; Venkatesh et al., 2003). But different groups of people have addressed the COVID-19 pandemic with different thinking, recognition, and behaviors. It will be valuable to investigate the way in which people behave heterogeneously; then, companies can accompany that information as they work to improve the performance of their workflow management system in satisfying their customers' requests. The relationship between effort expectancy and behavioral intention is vague. The moderating effects may strengthen the relationship between effort expectancy and behavioral intention.

The evaluation of DSRM used in this study was to assess the feasibility of the proposed digital resilience workflow management system by employing empirical tests on factors that impact customers' intention to adopt digital resilience, not on the productivity of the workflow management system. Future research could assess the effectiveness and productivity of ways to improve the entire workflow management system.

## MANAGERIAL IMPLICATIONS

This study seeks to construct reliable measures for organizations as they implement digital resilience and as they work to prevent their customers' contracting COVID-19 by helping the customers to avoid potential close physical contact. On the basis of the epidemiological SIR model, it is clear that close physical contact is a major reason why so many people became infected; in fact, it is the major reason why COVID-19 spread all over the world so rapidly. The avoidance of any potential close physical contact is an effective way to protect the susceptible from the infectious people. Although the authorities closed many local stores, grocery stores remain open for necessary daily needs. Digital resilience is the key measure that can assist a local store in implementing anti-COVID measures by setting up a contactless purchasing environment. In this way, potential close physical contact can be greatly reduced, and the store's customers will be safer.

Second, our study presents a workflow management system that solves a real problem: the avoidance of potential close physical contact in stores. The system could be a good example for companies that are seeking to facilitate their own digital resilience measures in order to mitigate the influences of COVID-19, especially in places with the potential for many people to gather, e.g., schools, hospitals, etc. The proposed workflow management system could easily be adjusted to fulfill the various standards and requirements of both organizations and individuals.

Third, the proposed workflow management system is a foundational framework, since it is clear that emerging technologies will be employed to improve organizations' digital capability. Many will look to implement the proposed workflow management system on a broad IoT (Internet of Things) platform integrated with both AI (artificial intelligence) and blockchain technology. IoT offers the potential to implement digital resilience to all of the devices within the system for information sharing, data storage, and performance estimation (Xu, He, & Li, 2014). AI could improve digital resilience by making the workflow management system more intelligent and automatic (Lu, 2019a), and blockchain technology offers a strong, decentralized platform that can provide security and privacy-preserving auditing for processing digital resilience (Lu, 2019b).

## CONCLUSIONS

A digital transformation has never been more urgently needed than it is now, following the unexpected disruptions from COVID-19. For a company to succeed in this world of unprecedented constraints upon its customers, it needs to empower enterprise information systems, to optimize operational activities, to foster the new culture of a hybrid work environment, and to engage its customers in new ways, intelligently and virtually transforming products and services with new business models. Digital resilience has the potential to help companies maintain their business performance and continuity in the COVID-19 world. Customers can adopt digital resilience to protect themselves from the threat of potential infection while completing necessary daily tasks. This study shows that customers are more willing to adopt digital resilience that is implemented by companies (e.g., grocery stores).

This study designs a digital resilience workflow management system that specifically focuses on protecting a business' customers from the infection of COVID-19 by assuring their avoidance of potential close physical contact with other shoppers. Another critical point is the customers' acceptance of digital resilience. Our findings demonstrate that the COVID-19 pandemic has forced customers to form new grocery shopping habits by using the digital resilience-enabled contactless method of grocery shopping. The institution of appropriate digital resilience-enabled measures is necessary a) to reduce the contact ratio ($q$) of COVID-19 and b) to keep customers both healthy and safe. It is expected that the more digital resilience-enabled companies will offer more competitive advantages that will both prevent the further dissemination of COVID-19 and will attract more customers for them, during the pandemic.

# APPENDIX SUPPLEMENTAL TABLES

### Table A1. Main Variables and Explanations

| SIR Model | | Workflow Management System | |
|---|---|---|---|
| Variable | Description | Variable | Description |
| $q$ | The contact ratio | $i$ | The starting place |
| $I$ | Infectious | $B$ | Business/Company |
| $I_0$ | The initial value of infectious | $C$ | Customer |
| $I_{MAX}$ | The maximal number of infectious | $F$ | A set of directed arcs |
| $R$ | Removed | $O_i$ | A terminal goal ($O_1, O_2, O_3$) |
| $R_0$ | The initial value of removal | $F_l$ | A labeling or weight function |
| $S$ | Susceptible | $P$ | A finite set of places. |
| $S_0$ | The initial value of susceptive | $T$ | A finite set of transitions |
| $TR$ | Total Population | $M_0$ | The start marking |
| $\alpha$ | The removal rate | $S_l$ | A finite set of activity labels |
| $\gamma$ | The rate of increase in the infectious | $\varphi$ | A set of messages |

### Table A2. Explanations of Messages between Customer and Business

| Message | Notification | Explanation |
|---|---|---|
| [(Access, C, B)] | Access represents access. | In (Access, C, B) means a store receives request of access from a customer. |
| [(Cap, B, C)] | Cap represents store capacity check. | Out (Cap, B, C) means a store sends capacity check to a customer. |
| [(Temp, SC, C)] | Temp represents customer body temperature check. | Out (Temp, SC, C) means Sensor Checking System sends temperature check to a customer. |
| [(Mask, SC, C)] | Mask represents customer wearing mask check. | Out (Mask, SC, C) means Sensor Checking System sends mask check message to a customer. |
| [(N_Cap, B, C)] | N_Cap represents a store isn't full. | In (N_Cap, B, C) means a store sends a message of it is not full to customer. |
| [(Y_Cap, B, C)] | Y_Cap represents a store is full. | In (Y_Cap, B, C) means a sore receives a message of it is full. |
| [(N_Tem, SC, C)] | N_Tem represents temperature check fails. | In (N_Tem, SC, C) means Sensor Checking System receives a message of temperature fails. Out (N_Tem, SC, C) means Sensor Checking System sends a message of temperature fails to a customer. |
| [(Y_Tem, SC, C)] | Y_Tem represents temperature check passes. | In (Y_Tem, SC, C) means Sensor Checking System receives a message of temperature passes from a store. |
| [(N_Mas, SC, C)] | N_Mas represents no wearing mask. | In (N_Mas, SC, C) means Sensor Checking System receives a message of mask check fails. Out (N_Mas, SC, C) means Sensor Checking System sends a message of maks check fails to a customer. |
| [(Y_Mas, SC, C)] | Y_Mas represents wearing mask. | In (Y_Mas, SC, C) means Sensor Checking System receives a message of mask check passes. |
| [(Pur, C, PM)] | Pur represents purchasing procedure. | In (Pur, C, PM) means Purchasing Monitoring System receives a message of purchasing from a customer. |
| [(Pur, PM, C)] | Pur represents purchasing procedure. | Out (Pur, PM, C) means Purchasing Monitoring System sends a message of purchasing to a customer. |
| [(Pay, C, PA)] | Pay represents payment procedure. | In (Pay, C, PA) means Payment Assistant System receives a message of payment from a customer. |
| [(Pay, PA, C)] | Pay represents payment procedure. | Out (Pay, PA, C) means Payment Assistant System sends a message of payment to a customer. |
| [(N_Deli, C, DA)] | N_Deli represents no delivery request. | Out (N_Deli, C, DA) means a customer sends a message of no delivery to Delivery Assistant System. |
| [(Y_Deli, C, DA)] | Y_Deli represents requesting delivery. | In [(Y_Deli, C, DA)] means Delivery Assistant System receives a message of delivery from a customer. |
| [(N_Ser, C, CS)] | N_Ser represents no customer service request. | Out [(N_Ser, C, CS)] means a customer sends a message of customer service to Customer Service System. |
| [(Y_Ser, C, CS)] | Y_Ser represents requesting customer service. | In [(Y_Ser, C, CS)] means Customer Service System receives a message of customer service from a customer. |
| Notes: <br> 1. The format of an exchanged message is: (Msg, Sender, Receiver). <br> 2. Msg is the key message, Sender or Receiver is one of the six role players. <br> 3. In represents a receiving message from sender to receiver, Out represents a sending message from sender to receiver. | | |

# APPENDIX 3. DSRM PROCESS MODEL